# On the Computing of the Minimum Distance of Linear Block Codes by Heuristic Methods

**Mohamed Askali, Ahmed Azouaoui, Saïd Nouh, Mostafa Belkasmi**
SIME Lab, National School of Computer Science and Systems Analysis (ENSIAS),
Mohammed V-Souisi University, Rabat, Morocco
Email: askali11@gmail.com

**ABSTRACT**

The evaluation of the minimum distance of linear block codes remains an open problem in coding theory, and it is not easy to determine its true value by classical methods, for this reason the problem has been solved in the literature with heuristic techniques such as genetic algorithms and local search algorithms. In this paper we propose two approaches to attack the hardness of this problem. The first approach is based on genetic algorithms and it yield to good results comparing to another work based also on genetic algorithms. The second approach is based on a new randomized algorithm which we call "Multiple Impulse Method (MIM)", where the principle is to search codewords locally around the all-zero codeword perturbed by a minimum level of noise, anticipating that the resultant nearest nonzero codewords will most likely contain the minimum Hamming-weight codeword whose Hamming weight is equal to the minimum distance of the linear code.

## 1. Introduction

The Minimum distance of a linear error correcting code has a practical and theoretical interest. It provides a great deal of information on the code capability in detecting and in correcting errors or erasures.

Since, to date, these problems cannot be solved mathematically because it is in general a NP-hard problem, it becomes necessary to physically search the codewords of a code in order to find the codeword with the minimum weight. Unfortunately, as the size of the code increases, the size of the search space becomes prohibitively large. The number of information bits in a code, k, defines the size of a search space. Note that k is also the number of basis vectors in the code and thus the size of the search space is 2 k. Thus, an exhaustive search is not feasible, but a heuristic search may provide valuable information and, in some cases, perhaps a solution. We propose several different algorithms and heuristic search techniques such as Genetic Algorithm (GA) [1-4], and search local error using a Soft-In decoder when applied to the problem of determining the true minimum distance of a linear block code [5].

In the past, many excellent studies have found the minimum weight for Quadratic Residue (QR) codes or its extended codes were presented in [6-10], in [11] we have estimated the minimum distance of Double Circulant Codes (DCC) using genetic algorithm, and Wallis *et al.* [12] have presented a different genetics techniques applied to find an estimate of the minimum distance for some Bose-Chaudhuri-Hocquenghem (BCH) codes, In [13], Nouh *et al.* have used genetic algorithms for finding a likelihood weight enumerator of some linear block codes, in particular their minimum weights.

Other works interest to the distance measurement methods have been introduced: Garello's true distance spectrum method [14], Berrou's error-impulse method [15], Garello's all-zero iterative decoding method [16] and Crozier's double (and triple) impulse method(s) [17].

Furthermore, there are also other works [18-20] based on artificial intelligence, trying to solve problems related to coding theory.

In this paper, we deal with finding a good estimate of minimum distance of linear block codes using genetic algorithms to BCH, QR, and DCC codes and which we denote $d_t$, and we compare our results to previous works. Finally, we present results obtained by using our search local error method published in a previous work [5], where we use a Soft-In Ordered Statistics decoder (OSD).

The remainder of this paper is organized as follow: in Section 2, we introduce the genetic algorithms; Section 3

describes the proposed heuristic methods to find a tight minimum distance, Section 4 reports the simulation results and discussions. Finally, Section 5 presents the conclusion and future trends.

## 2. Genetic Algorithms

Genetic Algorithms was first proposed by John Holland's, as a means to find good solutions to problems that were otherwise computationally intractable. Holland's schema theorem [21], and the related building block hypothesis, provided a theoretical and conceptual basis for the design of efficient GA. It also proved straight forward to implement GA due to their highly modular nature. As a consequence, the field grew quickly and the technique was successfully applied to a wide range of practical problems in science, engineering and industry. GA theory is an active and growing area, with a range of approaches being used to describe and explain phenomena not anticipated by earlier theory. In tandem with this, more sophisticated approaches for directing the evolution of a GA population are aimed at improving performance on classes of problem known to be difficult for GA, [21]. The development and success of GA contributed greatly to a wider interest in computational approaches based on natural phenomena. It is now a major stand of the wider field of computational intelligence, which encompasses techniques such as neural networks, and artificial immunology. Genetic algorithms are search methods that can be used for both solving problems and modelling evolutionary systems.

Since it is heuristic (it estimates a solution), GA differs from other heuristic methods in several ways. The most important difference is that it works on a population of possible solutions, while other heuristic methods use a Another important difference is that GA is not a deterministic but a probabilistic one.

A genetic algorithm is defined by (see **Figure 1**):

**Individual or chromosome:** a potential solution of the problem, it's a sequence of genes.

**Population:** a set of points of the research space.

**Environment:** the space of research.

**Fitness function:** the function to maximize/minimize.

**Encoding of chromosomes:** it depends on the treated problem, the famous known schemes of coding are: binary encoding, permutation encoding, value encoding and tree encoding.

**Stochastic Operators:**

- **Selection:** replicates the most successful solutions found in a population at a rate proportional to their relative quality.
- **Crossover:** Decomposes two distinct solutions and then randomly mixes their parts to form novel solutions.
- **Mutation:** Randomly perturbs a candidate solution. In the selection process, some individuals are selected to be copied into a tentative next population. Individual with higher fitness value is more likely to be selected. The selected individuals are altered by the mutation and crossover and form a new population of solutions. The GA is simple yet provides an adaptive and robust optimization methodology [22].

## 3. Estimation Methods for Finding the Minimum Distance

### 3.1. Methods Based on Genetic Algorithms

In the sequel of this paper, we use the following nota-

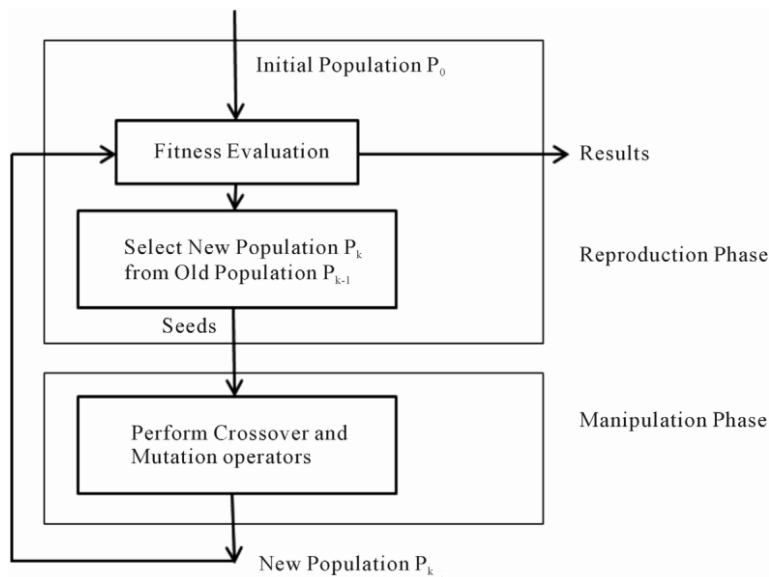

**Figure 1. The basic structure of the genetic algorithm.**

tions:
- $N_i$ the cardinal of the population.
- **Coding** the encoding function.
- $N_g$ the number of generations.
- $N_e$ the number of elites (better parents).
- $N_{gmax}$ is the maximum number of generations and C(I) is the codeword obtained by coding the information vector I.

In order to use genetic algorithms, in our work, we use binary encoding which consists to treat an individual as a binary sequence. We proposed Two GA variants A and B.

### 3.1.1. Genetic Algorithm: Variant A

This algorithm permits to find a minimum weight in a linear code C. It is known in field of coding theory that there exists always a linear systematic code equivalent to C. For the purposes of this paper we suppose that the generator matrix G of the code C is systematic; this chose permits to initialize the initial population by words of weight less than the global upper bound corresponding to the length n and the dimension k. the algorithm expects as inputs the probability of mutation $p_m$ of a single bit, and the crossover probability $p_c$.

- **Algorithm Steps**

The steps of the algorithm are organized as follow:

*Step 1: randomly generate an initial population*

Seed uniformly, randomly the initial population with a $N_i$, and where each individual is a word of length k with a random weight. We initiate the number of generation $N_g$ to 1.

*Step 2: while ($N_g < N_{gmax}$) do*

*Step 2.1: Compute the fitness of each individual in the population*

An individual i represents an information vector of k bits which is encoded by the code generator to an n-bit code vector. The fitness is the weight of the encoded individual if this last is different to zero otherwise, the fitness is equal to n as a maximum value.

$$f \leftarrow weight\bigl(Coding\bigl(individual\bigr)\bigr)$$

$$fitness\bigl(individual\bigr) = \begin{cases} f & \text{if } (f \neq 0) \\ n & \text{otherwise} \end{cases}$$

*Step 2.2: Sort population in increasing order of fitness*

*Step 2.3: Insert the best $N_e$ = 50% individuals in the intermediate population*

*Step 2.4: For i = $N_i$/2 to $N_i$*

*Step 2.4.1: Randomly select two individuals $p_1$ and $p_2$ for reproduction*

*Step 2.4.2: $p'_1$ = mutate ($p_1$) and $p'_2$ = mutate ($p_2$): Flip each bit of $p_1$ and $p_2$ with probability $p_m$*

*Step 2.4.3: Cross ( $p'_1$, $p'_2$ ) with probability $p_c$ to produce two children $ch_1$ and $ch_2$*

*Step 2.4.4: f1 ← weight (Coding (ch1)); f2 ← weight (Coding (ch2))*

*Step 2.4.5: if (f1 < f2) then insert $ch_1$ in the intermediate population else insert $h_2$*

*End For*

*End while*

*Step 3: output the first individual in the last population*

- **Description of the Algorithm**

In this entire paper, the crossover and the mutation stochastic operators operate only on the information bits represented as *k*-dimensional vectors. An alternative strategy is to represent individuals as *n* bit codewords.

In **Step 2.1**, to evaluate the fitness of an individual, it is necessary to first encode it by multiplying it with generator matrix G or by the generator polynomial if the code is cyclic as in some cases of our study. If the weight of the encoded vector is not null, the fitness is equal to weight (**Coding (vector)**) otherwise the fitness is equal to n. An individual is better than another if its weight is the smallest.

In **Step 2.2** and **Step 2.3**, we use a Linear Ranking Selection strategy where individuals in population are sorted by non-decreasing order of weight of encoded individual vector, and we select the best $N_e$ = 50% individuals to yield the intermediate population.

In **Step 2.4**, we use a single crossover point strategy, in which both parents organism strings is selected. All data beyond that point in either organism string is swapped between the two parents organisms. The resulting organisms are the children (see **Figure 2**).

Concerning selection, we use the random selection, in that only Ne individuals are preserved in the next generation, and we select randomly two parents to reproduce a best offspring that is more likely to contain good schema. The mutation step is done bit-wise on offspring with probability pm.

### 3.1.2. Genetic Algorithm: Variant B
- **Algorithm Steps**

The steps of the algorithm are organized as follow:

*Step 1: randomly generate an initial population*

Seed uniformly, randomly the initial population with a $N_i$, and where each individual is a word of length k with a random weight. We initiate the number of generation $N_g$ to 1.

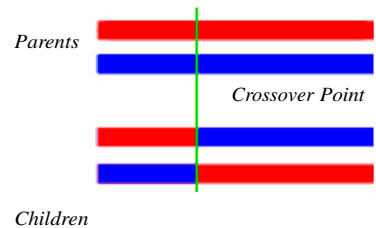

**Figure 2. The single crossover structure.**

*Step 2: while ($N_g < N_{gmax}$) do*
*Step 2.1: Compute the fitness of each individual in the population*

An individual *i* represents an information vector of *k* bits which is encoded by the generator code to an n-bit code vector. The fitness is the weight of the encoded individual if this last is different to zero otherwise, the fitness is equal to *n* as a maximum value.

$$f \leftarrow weight\left(Coding\left(individual\right)\right)$$

*Step 2.2: Sort population in increasing order of fitness*
*Step 2.3: select the best $N_e$ individuals in the intermediate population*
*Step 2.4: $i = N_e$ to $N_i$*
*Step 2.4.1: tournament select of two parents $p_1$ and $p_2$ for reproduction*
*Step 2.4.2: If (rand_value < $p_c$) {Cross $p_1$ and $p_2$ to generate $ch_1$ and $ch_2$; Mutate $ch_1$ and $ch_2$ and introduce them in the next population} Else introduce $p_1$ or $p_2$ into the next population with equal probability.*
*End For*
*Step 2.5: Let currbest = fittest of the intermediate population. If (fitness (best) < fitness (currbest)) best = currbest*
*End while*
*Step 3: output best*

- **Description of the Algorithm**

There are many differences between variant A and this variant in strategies of selection, order of stochastic operators, and the method of offspring reproduction.

In **Step 2.4.1**, we use the tournament selection, in that only one of two possible parents is preserved to reproduce two children whose will be inserted in the next generation.

**Step 2.4.2**, in this variant, the crossover operation depends on $p_c$, and it is done before the mutation step which is done bit-wise on offspring with probability $p_m$. In case of no-cross we insert the two initials parents in the next generation. We have used three strategies of crossover: a single crossover point (depicted in **Figure 2**), two point crossover, and uniform crossover. The two-point Crossover that randomly selects two crossover points within a chromosome then interchanges the two parent chromosomes between these points to produce two new offspring (see **Figure 3**). The Uniform Crossover uses a fixed mixing ratio between two parents. Unlike one- and two-point crossover, the Uniform Crossover enables the parent chromosomes to contribute the gene level rather than the segment level. An example of this operation is depicted in **Figure 4**.

### 3.2. A New Algorithm Based on Error Impulse Method

This method is not based on the analysis properties of the code but on the correction capability of the decoder. To obtain a good estimate of the minimum distance of a code, it is critical to apply a noise as possible to the all-zero codeword, so that the noise energy brings the decoder marginally away from the all-zero codeword.

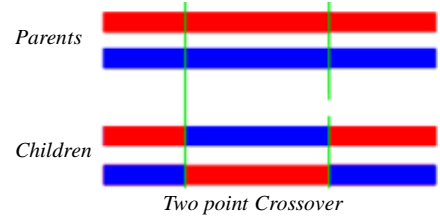

**Figure 3. Two-point crossover structure.**

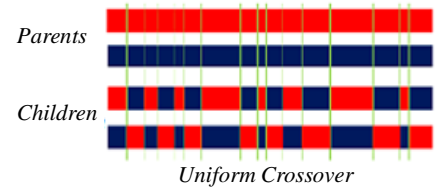

**Figure 4. Uniform crossover structure.**

The nature of the noise is so important, and it depends on the decoder used, for this, Berrou has proposed in [15] the Error impulse Method to excite the MLD decoder for turbo codes, and Xiao *et al.* [23] has proposed the Bit Reversing to excite the iterative decoder IRB proposed by Fossorier for LDPC codes. Another recent approach, by Garello *et al.* [24], is called the "all-zero iterative decoding algorithm". Here, this approach will be referred to as the "single impulse method" for reasons that will become apparent. This approach is similar to the error impulse method in that an impulse is again placed at a specific data index in the all-zero codeword. The main difference is that the amplitude of the impulse is intentionally set very high so that the decoder cannot correct it, but rather is forced to converge to (or at least select) a non-zero data pattern.

We denote by $X = (-1, -1-1, \cdots, -1, -1)$ the word associated to the "all-zero" codeword modulated by the BPSK.

#### 3.2.1. Berrou's Algorithm
The noise pattern proposed by Berrou *et al.* in [15] called Error Impulse, which was originally proposed for computing the minimum distance of Turbo codes.

In [15] C. Berrou developed and justified an algorithm with the MLD decoder to compute the minimum distance especially for Turbo codes, the principle of the idea is to send an "all-zero" codeword and adding gradually the level of noise and watch the capability of the decoder to return the "all-zero" codeword. Otherwise, this approach is based on inserting a (low-amplitude) error impulse into the all-zero codeword at a specific data index to see if the Turbo decoder can correct it. The amplitude of the error impulse is increased until the decoder fails. The highest amplitude that could be corrected provides an estimate of the minimum distance associated with that specific data index. An estimate of the overall minimum distance is obtained by testing all of the data indices in this manner. It was shown in [15] that this approach is guaranteed to find the true $d_t$ if the decoder is a true maximum likelyhood (ML) decoder. Of course, Turbo decoders are not true ML decoders. Thus, there is no guarantee that the error impulse method will find the true $d_t$. Further, although this approach is usually pessimistic, there is no guarantee that the result will be a true lower bound on $d_t$.

### 3.2.2. The Proposed Multiple Impulse Method (MIM)

The proposed algorithm produces a tight minimum distance based on true (low-weight) codewords found by a fine-tuned local search.

*We assume that $d_t$ is in the range $[d_0, d_1]$ where $d_0$ and $d_1$ are two integers. Then $d_t$ can be determined as follows*:

*Step 1: set Amin = $d_1$ + 0.5 and $d_t$ = n − $k_1$;*
*nb_test. Step 2: For i = 1 to nb_test*
*Step 2.1: A = $d_0$ − 0.5;*
*Step 2.2: Set [($\tilde{x} = x$) = TRUE];*
*Step 2.3: While [($\tilde{x} = x$) = TRUE] & [A ≤ Amin − 1.0]*
*Step 2.3.1: A = A + 1.0*
*Step 2.3.2: For nb_error = error_max to 1*
*Step 2.3.3: Subdivide A randomly on nb_error positions*
*Step 2.3.3.1: OSD decoding of Y → $\tilde{x}$*
*Step 2.3.3.2: If (weight ($\tilde{x}$) ≤ $d_t$) then dt = weight ($\tilde{x}$)*
*Step 2.3.3.3: If ($\tilde{x} = x$) then [($\tilde{x} = x$) = TRUE]*
*End for*
*End while*
*Step 2.4: Amin = A;*
*End for*
*Output: $d_t$ is the minimum distance*

We changed the Soft-In MLD decoder used in Berrou's [15] Algorithm by a Soft-In OSD decoder which is very fast, by injecting a noise iteratively in a random positions, the decoder word will be mostly near than the "all-zero" codeword, and the minimum distance of the code will be the minimum weight of the decoded words and is not the magnitude $A_i^*$ of the noise as we have in Error Impulse Method.

## 4. Results and Discussions

### 4.1. Parameters Optimization of Genetic Algorithms

In **Tables 1-4**, we analyze the impact of genetic operators on the minimum distance.

**Table 1. Effect of elitism operator.**

| Codes | Minimum distance | |
|---|---|---|
| | Without elitism | With elitism |
| BCH (255, 99, 47) | 58 | 52 |
| BCH (255, 107, 45) | 53 | 53 |
| BCH (255, 115, 43) | 46 | 46 |
| QR (193, 97, 27) | 28 | 27 |
| QR (199, 100, 31) | 35 | 31 |
| QR (223, 112, 31) | 36 | 32 |

**Table 2. Effect of crossover operator types.**

| Codes | Minimum distance | | |
|---|---|---|---|
| | 1-point | 2-point | Uniform |
| BCH (255, 99, 47) | 58 | 52 | 57 |
| BCH (255, 107, 45) | 51 | 53 | 52 |
| BCH (255, 115, 43) | 48 | 46 | 46 |
| QR (193, 97, 27) | 30 | 27 | 30 |
| QR (199, 100, 31) | 31 | 31 | 32 |
| QR (223, 112, 31) | 35 | 32 | 35 |

**Table 3. Effect of selection types.**

| Codes | Minimum distance | | |
|---|---|---|---|
| | Tournament | Random | Roulette |
| BCH (255, 99, 47) | 52 | 53 | 53 |
| BCH (255, 107, 45) | 53 | 50 | 53 |
| BCH (255, 115, 43) | 46 | 46 | 46 |
| QR (193, 97, 27) | 27 | 30 | 30 |
| QR (199, 100, 31) | 31 | 31 | 31 |
| QR (223, 112, 31) | 32 | 35 | 32 |

**Table 4. Effect of mutation operator types.**

| Codes | Minimum distance | |
|---|---|---|
| | Classic mutation | Non classic mutation |
| BCH (255, 99, 47) | 52 | 57 |
| BCH (255, 107, 45) | 53 | 57 |
| BCH (255, 115, 43) | 46 | 48 |
| QR (193, 97, 27) | 27 | 32 |
| QR (199, 100, 31) | 31 | 35 |
| QR (223, 112, 31) | 32 | 40 |

### 4.1.1. Effect of Elitism
It appears that elitism significantly improves the performance of genetic algorithm for QR codes and BCH codes.

### 4.1.2. Effect of Crossover
These results explain that 2-point crossover seems to perform significantly better than uniform and 1-point crossover for QR codes and BCH codes.

### 4.1.3. Effect of Selection
As shown results in the **Table 3**, in general, the tournament selection gives a very close upper bound to the minimum distance for QR codes and BCH codes.

### 4.1.4. Effect of Mutation
In this paragraph, we compare the impact of the classic mutation with other type of mutation. This last alters an individual by bit inversion of chromosome. However, such an inversion takes place only in one bit and only when an improvement in the individual's fitness is achieved. If it is not possible to improve the individual's fitness, then no alteration is performed. The algorithm simply goes through every chromosome's gene to determine which of them must be changed in such a way as to improve individual's fitness.

According to the results of this study, we concluded that the best parameters for this algorithm are: Elitism operator, tournament selection, 2-point crossover, and classical mutation.

## 4.2. Results of Various Genetic Algorithms

All simulations were made with default GA parameters outlined in the **Table 5**.

The **Table 6** shows that the proposed algorithms out-performed the other optimization techniques.

**Table 5. Parameters of implementation of genetic algorithms.**

| Parameter | Wallis's GA | GA-A | GA-B |
|---|---|---|---|
| Probability of Crossover | 80% | 93% | 80% |
| Probability of Mutation | 2% | 1% | 2% |
| Crossover Type | 2-point | 1-point | 2-point |
| Selection Type | Tournament | Random | Tournament |
| Tournament Size | 3 | - | 2 |
| Generation Number | 75 | 75 | 75 |
| Individuals Number | 10,000/1000 | 10,000/1000 | 10,000/1000 |

**Table 6. Comparaison of our GA algorithms with other works for some BCH codes.**

| Codes BCH (n, k, *d*-design) | $d_t$ GA-A 10000 | $d_t$ GA-B 10000 | Wallis's GA | Hill-Climbing | Tabu Search |
|---|---|---|---|---|---|
| BCH (127, 64, 21) | 21 | 21 | 21 | 28 | 24 |
| BCH (127, 57, 23) | 23 | 23 | 23 | 28 | 23 |
| BCH (127, 50, 27) | 27 | 27 | 27 | 32 | 31 |
| BCH (255, 71, 59) | 64 | 63 | 66 | 79 | 79 |
| BCH (255, 79, 55) | 57 | 57 | 60 | 74 | 64 |
| BCH (255, 87, 53) | 57 | 58 | 57 | 70 | 66 |
| BCH (255, 91, 51) | 58 | 53 | 59 | 72 | 69 |
| BCH (255, 99, 47) | 51 | 52 | 55 | 64 | 61 |
| BCH (255, 107, 45) | 53 | 49 | 51 | 64 | 62 |
| BCH (255, 115, 43) | 48 | 45 | 50 | 57 | 55 |
| BCH (511, 304, 51) | 87 | 74 | 79 | 90 | 85 |
| BCH (511, 286, 55) | 98 | 84 | 84 | 96 | 92 |
| BCH (511, 238, 75) | 113 | 103 | 105 | 118 | 112 |
| BCH (511, 220, 79) | 112 | 109 | 111 | 123 | 117 |
| BCH (511, 184, 91) | 111 | 127 | 128 | 135 | 140 |
| BCH (511, 166, 95) | 143 | 135 | 137 | 152 | 140 |
| BCH (511, 121, 117) | 159 | 155 | 152 | 163 | 163 |
| BCH (511, 103, 123) | 164 | 160 | 164 | 179 | 179 |
| BCH (511, 76, 171) | 176 | 176 | 176 | 195 | 184 |
| BCH (511, 58, 183) | 183 | 184 | 185 | 207 | 199 |

The **Table 7** shows the computational results of minimum distance via GA-A, GA-B, and the Simulated Annealing (SA) developed by authors in [25].

In the **Table 7**, for the three first BCH codes listed, these algorithms found the true minimum distance. However, for the two last BCH codes, the gap between the minimum distance obtained by the SA algorithm and the true value is still large, while our genetic algorithms found this true minimum distance.

The **Table 8** shows that our two variants of GA give the same estimate of the minimum distance for Quadratic residue codes where the length is less than 223.

The **Table 9**, we validate our estimate minimum distance by the exhaustive method for some random DCC defined by their binary header.

**Table 7. Comparaison between our two GA variants with simulated annealing.**

| Codes BCH (n, k, *d*-design) | $d_t$ GA-A 10000 | $d_t$ GA-B 10000 | Simulated annealing |
|---|---|---|---|
| BCH (15, 11, 3) | 3 | 3 | 3 |
| BCH (31, 26, 3) | 3 | 3 | 3 |
| BCH (63, 24, 15) | 15 | 15 | 15 |
| BCH (127, 64, 21) | 21 | 21 | 27 |
| BCH (255, 91, 51) | 51 | 51 | 75 |

**Table 8. Comparaison between our two variants applied to QR codes.**

| Codes QR (n, k, d) | $d_t$ GA-A 1000 | $d_t$ GA-B 1000 |
|---|---|---|
| QR (47, 24, 11) | 11 | 11 |
| QR (71, 36, 11) | 11 | 11 |
| QR (73, 37, 13) | 13 | 13 |
| QR (79, 40, 15) | 15 | 15 |
| QR (89, 45, 17) | 17 | 17 |
| QR (97, 49, 15) | 15 | 15 |
| QR (113, 57, 15) | 15 | 15 |
| QR (127, 64, 19) | 19 | 19 |
| QR (137, 69, 21) | 21 | 21 |
| QR (151, 76, 19) | 19 | 19 |
| QR (191, 96, 27) | 27 | 27 |
| QR (193, 97, 27) | 30 | 30 |
| QR (199, 100, 31) | 31 | 31 |
| QR (223, 112, 31) | 32 | 32 |

### 4.3. Validation and Results of Multiple Impulse Method

#### 4.3.1. Validation of the Multiple Impulse Method

All simulations have been done using a simple configuration machine: Intel® Core™ 2 CPU T5600 @ 1.83 GHz, RAM: 2.00 GHz.

As a first step, we validated the algorithm by verifying the minimum distance for some linear codes: BCH codes, Quadratic residue Codes, and Quadratic Double-Circulant Codes in their Bordered form, for which the minimum distance is known.

The results are summarized in the **Tables 10-12**, in which "OSD_EI" denotes the Order Statistic Decoding with Error impulse and "TTE" denotes the Time of execution in seconds. As it is shown in these tables our algorithm is successfully validated.

Let p be a prime that is congruent to ±3 modulo 8. A binary [2(p + 1), p + 1] quadratic double-circulant code (QDC) [26], denoted by B, can be constructed using the following defining polynomials:

$$b(x) = \begin{cases} 1+\sum_{r \in Q} x^r & \text{if } p \equiv 3 \pmod{8} \text{ and} \\ \sum_{r \in Q} x^r & \text{if } p \equiv -3 \pmod{8} \end{cases} \quad (1)$$

Q is the set of quadratic residues modulo p.

The generator matrix G of B can be written as described in **Figure 5**.

We find exactly what Tomlinson has found in [26].

#### 4.3.2. New Experimental Results

In this section we present an application of the proposed algorithm to find the true unknown minimum distance of some residue quadratic codes (QR and QDC see **Tables 14-16**) and some BCH codes (see **Table 13**) comparing respectively to some known upper bounds, and to the designed distance or comparing to the Grassl table [27].

By MacWilliams and Sloane in [28], for QR codes, we compare our estimation by this inequality $d^2 \geq n$.

By the Pless's identity [29], the minimum weight in Quadratic Residue codes is always odd. This means that when we find a codeword with a pair weight w, it is necessary to have a codeword with an odd weight w-1.

In the **Table 14** we give some QR codes where the length is like the form **n = 8 m + 1**.

In our knowledge the Krasikov Bound [30] is the best upper bound for comparing our estimation of the minimum distance. In the **Table 15** we give some QR codes

$$G = \begin{bmatrix} 1 & & & 0 & & \\ \vdots & I_p & & \vdots & B & \\ 1 & & & 0 & & \\ 0 & 0 & \cdots & 0 & 1 & 1 & \cdots & 1 \end{bmatrix}$$

**Figure 5. The quadratic double circulant form.**

**Table 9. Comparaison between our two variants applied to QR codes.**

| Double Circulant Codes (DCC) | Binary Header of DCC | $d_t$ Our GA-A 1000 Individuals | $d_t$ Our GA-B 1000 Individuals | Exhaustive Method |
|---|---|---|---|---|
| C (20, 10) | 1001111110 | 6 | 6 | 6 |
| C (22, 11) | 00010110111 | 7 | 7 | 7 |
| C (24, 12) | 101000110111 | 8 | 8 | 8 |
| C (26, 13) | 1000100111100 | 7 | 7 | 7 |
| C (28, 14) | 00101001111111 | 8 | 8 | 8 |
| C (30, 15) | 001110111111101 | 8 | 8 | 8 |
| C (32, 16) | 1010100100100110 | 8 | 8 | 8 |
| C (34, 17) | 10011001011010011 | 8 | 8 | 8 |
| C (36, 18) | 101000100011111111 | 8 | 8 | 8 |
| C (38, 19) | 1100100000111111101 | 8 | 8 | 8 |
| C (40, 20) | 00011100110011010101011 | 9 | 9 | 9 |
| C (42, 21) | 000101111011110011110 | 10 | 10 | 10 |
| C (44, 22) | 1100011101010101001111 | 10 | 10 | 10 |
| C (46, 23) | 01101101111101011110000 | 11 | 11 | 11 |
| C (50, 25) | 1001000111111001011000000 | 10 | 10 | 10 |
| C (52, 26) | 11000100110110001110110010 | 10 | 10 | 10 |
| C (54, 27) | 011000110000111111101101000 | 11 | 11 | 11 |
| C (58, 29) | 00011011111000110010010010010 | 12 | 12 | 12 |
| C (62, 31) | 1100001010100011100000011010110 | 12 | 12 | 12 |

**Table 10. Validation of some QR codes with known mini- mum distance.**

| Codes QR (n, k, d) | $d_t$ OSD_EI | TTEs |
|---|---|---|
| QR (41, 21, 9) | 9 | 1.12 |
| QR (47, 24, 11) | 11 | 0.61 |
| QR (71, 36, 11) | 11 | 0.71 |
| QR (73, 37, 13) | 13 | 0.75 |
| QR (79, 40, 15) | 15 | 1.29 |
| QR (89, 45, 17) | 17 | 3.66 |
| QR (97, 49, 15) | 15 | 0.99 |
| QR (113, 57, 15) | 15 | 4.23 |
| QR (127, 64, 19) | 19 | 11.08 |
| QR (137, 69, 21) | 21 | 14.33 |
| QR (151, 76, 19) | 19 | 33.01 |
| QR (191, 96, 27) | 27 | 213 |
| QR (193, 97, 27) | 27 | 220 |
| QR (199, 99, 31) | 31 | 145 |
| QR (223, 112, 31) | 31 | 124 |

where the length is like the form **n = 8m − 1**, and we have: $d \leq 0.166315\, n$.

For the class of QDC codes we give in the next table some codes with unknown minimum.

**Table 11. Validation of some BCH codes with minimum distance.**

| Codes BCH (n, k, d-design) | $d_t$ OSD_EI | TTEs |
|---|---|---|
| BCH (31, 16, 7) | 7 | 0.02 |
| BCH (31, 21, 5) | 5 | 0.01 |
| BCH (63, 18, 21) | 21 | 0.04 |
| BCH (63, 24, 15) | 15 | 0.06 |
| BCH (63, 36, 11) | 11 | 0.07 |
| BCH (63, 39, 9) | 9 | 0.05 |
| BCH (63, 30, 13) | 13 | 0.04 |
| BCH (63, 45, 7) | 7 | 0.06 |
| BCH (63, 51, 5) | 5 | 0.05 |
| BCH (63, 57, 3) | 3 | 0.01 |
| BCH (127, 8, 63) | 63 | 0.67 |
| BCH (127, 15, 55) | 55 | 0.7 |
| BCH (127, 22, 47) | 47 | 0.37 |
| BCH (127, 29, 43) | 43 | 0.51 |
| BCH (127, 71, 19) | 19 | 2.85 |
| BCH (127, 78, 15) | 15 | 2.08 |
| BCH (127, 92, 11) | 11 | 1.37 |
| BCH (127, 106, 7) | 7 | 0.74 |
| BCH (255, 45, 87) | 87 | 57 |
| BCH (255, 55, 63) | 63 | 49 |

**Table 12. Validation of some QDC codes with known minimum distance.**

| Codes QDC (2(p + 1), p + 1, d) | $d_t$ OSD_EI | TTEs |
|---|---|---|
| QDC (24, 12, 8) | 8 | 0.97 |
| QDC (28, 14, 8) | 8 | 1.02 |
| QDC (40, 20, 8) | 8 | 1.4 |
| QDC (60, 30, 12) | 12 | 2.86 |
| QDC (76, 38, 12) | 12 | 6.61 |
| QDC (88, 44, 16) | 16 | 9.88 |
| QDC (108, 54, 20) | 20 | 34.98 |
| QDC (120, 60, 20) | 20 | 33.33 |
| QDC (124, 62, 20) | 20 | 33.18 |
| QDC (136, 68, 24) | 24 | 57 |
| QDC (168, 84, 24) | 24 | 1052 |

**Table 13. Tight bound for unknown minimum distance of some BCH codes.**

| Codes BCH (n, k, d-design) | $d_t$ OSD_EI | TTEs |
|---|---|---|
| BCH (127, 64, 21) | 21 | 10 |
| BCH (127, 57, 23) | 23 | 1 |
| BCH (127, 50, 27) | 27 | 4 |
| BCH (255, 71, 59) | 62 | 530 |
| BCH (255, 79, 55) | 55 | 631 |
| BCH (255, 87, 53) | 53 | 5915 |
| BCH (255, 91, 51) | 51 | 7617 |
| BCH (255, 115, 43) | 43 | 8283 |
| BCH (255, 123, 39) | 39 | 7098 |
| BCH (255, 131, 37) | 37 | 7570 |
| BCH (255, 139, 31) | 31 | 7051 |
| BCH (255, 147, 29) | 29 | 4626 |
| BCH (255, 155, 27) | 27 | 4177 |
| BCH (255, 163, 25) | 25 | 2612 |
| BCH (255, 171, 23) | 23 | 2847 |
| BCH (255, 179, 21) | 21 | 1653 |
| BCH (255, 187, 19) | 19 | 65 |
| BCH (255, 191, 17) | 17 | 1198 |
| BCH (255, 199, 15) | 15 | 384 |

**Table 14. Tight bound of the unknown minimum distance of some QR codes.**

| Codes QR | | Square (n) | $d_t$ OSD_EI | TTEs |
|---|---|---|---|---|
| n = 233 | k = 117 | 15.26 | 25 | 198 |
| n = 241 | k = 121 | 15.52 | 31 | 835 |
| n = 257 | k = 129 | 16.03 | 33 | 5851 |
| n = 281 | k = 141 | 16.76 | 35 | 40435 |
| n = 313 | k = 157 | 17.69 | 45 | 51498 |
| n = 337 | k = 169 | 18.35 | 51 | 45539 |

**Table 15. Tight bound of the unknown minimum distance of some QR codes.**

| Codes QR | | Krasikov bound | $d_t$ OSD_EI | TTEs |
|---|---|---|---|---|
| n = 239 | k = 129 | 39.74 | 31 | 308 |
| n = 263 | k = 132 | 43.74 | 35 | 66173 |
| n = 271 | k = 136 | 45.07 | 39 | 65659 |
| n = 311 | k = 156 | 51.72 | 35 | 62286 |
| n = 359 | k = 180 | 59.70 | 55 | 60621 |
| n = 367 | k = 184 | 61.03 | 59 | 74650 |
| n = 383 | k = 192 | 63.69 | 59 | 81810 |
| n = 431 | k = 216 | 71.68 | 67 | 101409 |
| n = 439 | k = 220 | 73.01 | 71 | 90579 |

**Table 16. Tight bound of the unknown minimum distance of some QDC codes.**

| Codes QDC (2(p + 1), p + 1) | $d_t$ OSD_EI | TTEs |
|---|---|---|
| QDC (204, 102) | 24 | 5886 |
| QDC (216, 108) | 24 | 1421 |
| QDC (220, 110) | 30 | 7595 |
| QDC (264, 132) | 40 | 27547 |
| QDC (280, 140) | 36 | 24683 |
| QDC (300, 150) | 44 | 28371 |
| QDC (316, 158) | 46 | 35892 |
| QDC (328, 164) | 48 | 53135 |

## 5. Conclusion and Perspectives

In this paper we have used genetic algorithms to find a good estimate of minimum distance for BCH, DCC, and QR codes. The implementation of proposed genetic algorithms shows that they are more efficient comparing to competitor genetic algorithm developed by Wallis.

For the same goal, we have proposed the Multiple Impulse Method based on Soft-In OSD decoding algorithm by generalization of the method proposed initially by Berrou *et al.* The MIM technique is highly performing as a good tool for computing the minimum distance of linear codes, especially for a large code where the length is so long. In the perspectives of this work, we have to apply these powerful tools to construct good linear block codes, and to test the effect of other Soft-In decoders in terms of complexity and performances.